\documentclass[iop]{emulateapj}

\begin{document}

\title{H$_2$CO distribution and formation in the TW Hya disk}

\author{Karin I. \"Oberg, Viviana V. Guzm{\'a}n\altaffilmark{1}, Christopher J. Merchantz\altaffilmark{2}, Chunhua Qi, Sean M. Andrews, L. Ilsedore Cleeves, Jane Huang, Ryan A. Loomis, David J. Wilner}
\affil{Harvard-Smithsonian Center for Astrophysics, 60 Garden St., Cambridge, MA 02138}


\author{Christian Brinch}
\affil{Niels Bohr International Academy, The Niels Bohr Institute, University of Copenhagen, Blegdamsvej 17, DK-2100 Copenhagen {\O}, Denmark
}


\author{Michiel Hogerheijde}
\affil{Leiden Observatory, Leiden University, PO Box 9513, 2300 RA, Leiden, the Netherlands}


\begin{abstract}

H$_2$CO is one of the most readily detected organic molecules in
protoplanetary disks. Yet its distribution and dominant formation pathway(s)
remain largely unconstrained. To address these issues, we present ALMA observations of two H$_2$CO lines ($3_{12}-2_{11}$ and $5_{15}-4_{14}$) at 0$\farcs$5 ($\sim$30~au) spatial resolution  toward the disk around the nearby T Tauri star TW Hya. Emission from both lines is spatially resolved, showing a central depression, a peak at $0\farcs4$ radius, and a radial
decline at larger radii with a bump at $\sim1''$, near the millimeter
continuum edge. We adopt a physical model for the disk and use toy models
to explore the radial and vertical H$_2$CO abundance structure. We find that
the observed emission implies the presence of at least two distinct H$_2$CO
gas reservoirs: (1) a warm and unresolved inner component ($<10$ au), and
(2) an outer component that extends from $\sim15$ au to beyond the millimeter
continuum edge. The outer component is further constrained by the line ratio
to arise in a more elevated disk layer at larger radii. The inferred H$_2$CO
abundance structure agrees well with disk chemistry models, which
predict efficient H$_2$CO gas-phase formation close to the star, and cold H$_2$CO grain
surface formation, through H additions to condensed CO, followed by non-thermal desorption in the outer disk.
The implied presence of active grain surface chemistry in the TW Hya disk is consistent
with the recent detection of CH$_3$OH emission, and suggests that more complex
organic molecules are formed in disks, as well.

\end{abstract}


\section{Introduction} \label{sec:intro}

Planets are assembled and obtain their initial organic compositions from solids and gas in protoplanetary disks. Terrestrial, rocky planets are expected to form close to their stars and directly sample the inner disk refractory organics, though some volatile organics could be added through direct accretion of disk gas. More volatile organic material from the outer disk can become incorporated into the planet by later planetesimal bombardment \citep{Morbidelli12,Raymond14}. The abundance of volatile organics on nascent planets are of particular interest, since they can drive a complex prebiotic chemistry leading to formation of the different building blocks of RNA and proteins \citep{Powner09}. 

Based on cometary studies, volatile organics were common in the Solar Nebula; comets frequently contain anywhere between a few and 10\% of volatile organics with respect to water ice \citep{Mumma11,LeRoy15}. The most abundant ones are CH$_3$OH, CH$_4$, C$_2$H$_2$, H$_2$CO, C$_2$H$_6$ and HCN. Of these, CH$_3$OH, C$_2$H$_2$, H$_2$CO and HCN have also been detected in gas form in protoplanetary disks, suggesting that, similar to the Solar System planets, exoplanets form in environments rich in volatile organic species \citep[e.g.][]{Dutrey97,Aikawa03,Carr08,Walsh16}. 

Of these molecules, H$_2$CO and CH$_3$OH are of special prebiotic interest. Both can form through grain surface hydrogenation of condensed CO and become incorporated into icy bodies. Based on laboratory experiments, such organic-rich ices become sources of a range of complex organic molecules when exposed to any kind of high-energy radiation or electrons \citep[e.g.][]{Gerakines96,Hudson00,Bennett07a,Oberg09d,Oberg16,Boyer16,Sullivan16}. CH$_3$OH only forms through ice chemistry, and if it was readily observable it would be the best tracer of organic ice chemistry in disks. CH$_3$OH is challenging to detect, however, due to its low volatility and large partition function. To date it has only been observed in a single disk at low SNR, resulting in very limited constraints on its radial or vertical distribution \citep{Walsh16}. 

H$_2$CO is easier to observe \citep{Aikawa03,Oberg10c,Oberg11a,vanderMarel14}, but connecting these observations to disk ice chemistry is complicated by its viable gas-phase formation pathways. High spatial resolution observations are needed to decide between gas and grain surface formation pathways. H$_2$CO grain surface formation would only be expected where it is cold enough for CO to accrete onto grains and remain there for a sufficient time to allow chemical reactions with H \citep{Watanabe02,Fuchs09,Cuppen09}. In a disk with a radially decreasing temperature profile, H$_2$CO formed through such grain surface chemistry should only appear at a distance from the central star corresponding to midplane temperatures below 20--30~K \citep{Fayolle16}. Though possible everywhere in the disk, gas-phase H$_2$CO formation is expected to occur most efficiently in the warm and dense inner disk, producing a centrally peaked H$_2$CO abundance and emission profile. 

H$_2$CO has  been observed at high spatial resolution with ALMA in one disk, around DM Tau; where \citet{Loomis15} found that H$_2$CO is distributed throughout the disk, i.e a hybrid of what is expected from pure gas-phase or grain-surface chemistry. This  result was used to conclude that H$_2$CO forms through both gas and grain surface chemistry in this disk, since neither pathway can explain all the observed emission. 

In this study, we revisit the distribution and chemistry of H$_2$CO in disks by characterizing its abundance pattern in an older example, the disk around TW Hya. Because TW Hya is nearby (d=59~pc \citep{GAIA16}), analogous observations provide access to smaller physical scales compared to DM Tau, potentially providing a clearer separation between H$_2$CO disk components originating through gas and grain surface chemistry. TW Hya is also a good target to interpret observed H$_2$CO abundance patterns, since it is a well characterized protoplanetary disk both in terms of physical structure \citep[e.g.][]{Bergin15,Andrews16,Schwarz16} and chemistry \citep[e.g.][]{Kastner97,Thi04,Walsh16}, including constraints on the CO snowline location \citep{Qi13c}. 

We present 0$\farcs$5 resolution ALMA Cycle 2 observations of two H$_2$CO lines toward the TW Hya protoplanetary disk: H$_2$CO $3_{12}-2_{11}$ and $5_{15}-4_{14}$. \S2 describes the observations and presents the observed H$_2$CO emission. In \S3 we present a series of toy models of different H$_2$CO distributions, and compare the model output with observations  to constrain the H$_2$CO abundance profile. In \S4 we discuss the distribution of H$_2$CO in the TW Hya disk, its connections to known physical and chemical structures, and implications for the formation chemistry of H$_2$CO (and other organics) during planet formation. \S5 presents some concluding remarks.

\section{Observations} \label{sec:obs}

\subsection{Observational Details}

This paper makes use of ALMA Cycle~2 observations of two different H$_2$CO lines toward the young star TW Hya. H$_2$CO $3_{12}-2_{11}$ was observed  on 2014 July 19 as a part of  ADS/JAO.ALMA\#2013.1.00114.S (PI: K. \"Oberg) with 31 antennas and baselines ranging from 30 to 650~m. H$_2$CO $5_{15}-4_{14}$ was observed on December~31, 2014 and June~15, 2015 with 34 antennas ($15-349$~meter baselines) and 36 antennas ($21-784$~meter baselines), respectively as a part of ADS/JAO.ALMA\#2013.1.00198.S (PI: E. Bergin).

For the 2014 July observations, the quasar J1037-2934 was used for both bandpass and phase calibration, and Pallas for flux calibration.  
The H$_2$CO $3_{12}-2_{11}$  transition (Table~\ref{lines}) was observed with a channel width of 122kHz ($\sim$0.16 km/s).  The total on-source integration time was 41~minutes.  Prior to imaging, the pipeline-calibrated data from JAO were phase and amplitude self calibrated on the continuum in the H$_2$CO spectral window using CASA version 4.5 and timescales of 10--30~seconds.  This increased the SNR of the emission by a factor of $\approx3$.
The line data were continuum subtracted and imaged. We CLEANed \citep{Hogbom74} the images with a 0.25 km/s resolution down to a level of $3\times$rms. During the CLEANing process we employed a mask, constructed by manually identifying areas with emission in each channel, and Briggs parameter of 0.5. We used a separate line-free spectral window with a frequency width of 469 MHz to generate a continuum image. The total continuum flux is 560$\pm$84~mJy, assuming a 15\% absolute flux calibration uncertainty. This is consistent with the previously measured flux of 540~mJy with the Submillimeter Array for a similar frequency range \citep{Qi06}.

For the 2015 June and 2014 December H$_2$CO $5_{15}-4_{14}$ observations, the quasars J1256-057 and J1037-2934 were used for bandpass and gain calibration, respectively. Titan was used for the flux calibration. The H$_2$CO $5_{15}-4_{14}$ transition was observed using a channel width of 244kHz (0.21 km/s).  The total on-source integration time for was 43~minutes.  There was a pointing misalignment that may in part be due to TW Hya's high proper motion, and so we aligned phase-centers of the compact and extended data sets based on the continuum peak location \citep{Bergin16}. The data were then self-calibrated, CLEANed, and imaged similarly to the H$_2$CO $3_{12}-2_{11}$ data. 

The resulting H$_2$CO line peak and disk integrated fluxes are reported in Table~\ref{lines} with rms uncertainties. An additional 15\% uncertainty should be applied to account for the absolute flux calibration uncertainty. 

\begin{figure*}[ht!]
\figurenum{1}
\plotone{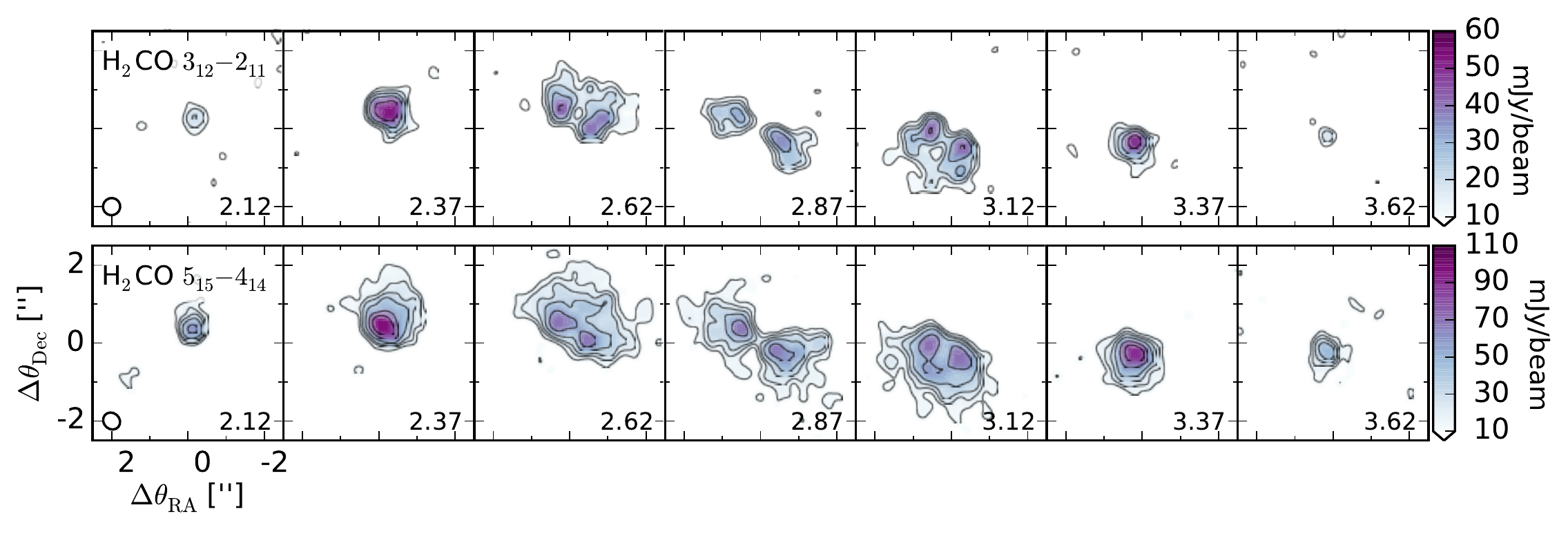}
\caption{Channel maps of  of H$_2$CO $3_{12}-2_{11}$ (top), H$_2$CO $5_{15}-4_{14}$ (bottom) with 0.25 km/s channels. The flux per beam is indicated by the color scales. The line contours are [3,5,7,10,15,20]$\sigma$.\label{fig:ch_map}}
\end{figure*}

\begin{deluxetable*}{lcccccc}
\tablecaption{Observational data\label{lines}}
\tablehead{
\colhead{Line}   & \colhead{Rest freq.}    & \colhead{Log$\rm_{10}(A_{ij})$	} &
\colhead{E$_{\rm u}$}    & \colhead{beam (PA)}	 & \colhead{Peak integrated flux} & \colhead{Peak flux\tablenotemark{a}}\\
 \colhead{}   & \colhead{GHz}    & \colhead{} &
\colhead{K}    &\colhead{$\arcsec\times\arcsec$ ($^{\circ}$)} &\colhead{mJy km/s beam$^{-1}$}& \colhead{mJy beam$^{-1}$}
}
\startdata
H$_2$CO	$3_{12}-2_{11}$	&225.69778	&-3.56	&33.4	&$0\farcs45\times0\farcs45$ ($-75^{\circ}$)	&26.7$\pm$2.5	&52.8$\pm$2.9\\
H$_2$CO	$5_{15}-4_{14}$	&351.76864	&-2.43	&62.5	&$0\farcs47\times0\farcs41$ ($51^{\circ}$)	&57.0$\pm$3.9	&96.3$\pm$3.8\\
\enddata
\tablenotetext{a}{in 0.25 km/s channels}
\end{deluxetable*}

\subsection{H$_2$CO Spectral Image Cubes}

Figure \ref{fig:ch_map} presents channel maps of H$_2$CO $3_{12}-2_{11}$ and H$_2$CO	$5_{15}-4_{14}$ toward the TW Hya protoplanetary disk. The data was resampled to place the central channel at 2.87~km/s -- close to the previously observed systemic velocity of TW Hya \citep{Hughes11}. Both lines display clear rotation patterns, consistent with a Keplerian disk. The $5_{15}-4_{14}$ emission is more extended than the $3_{12}-2_{11}$ emission, which may be partially a sensitivity issue -- the rms noise in the $5_{15}-4_{14}$ data is $\sim$50\% higher than in the $3_{12}-2_{11}$ data, but the $5_{15}-4_{14}$ transition is intrinsically an order of magnitude stronger.

\begin{figure*}
\figurenum{2}
\plotone{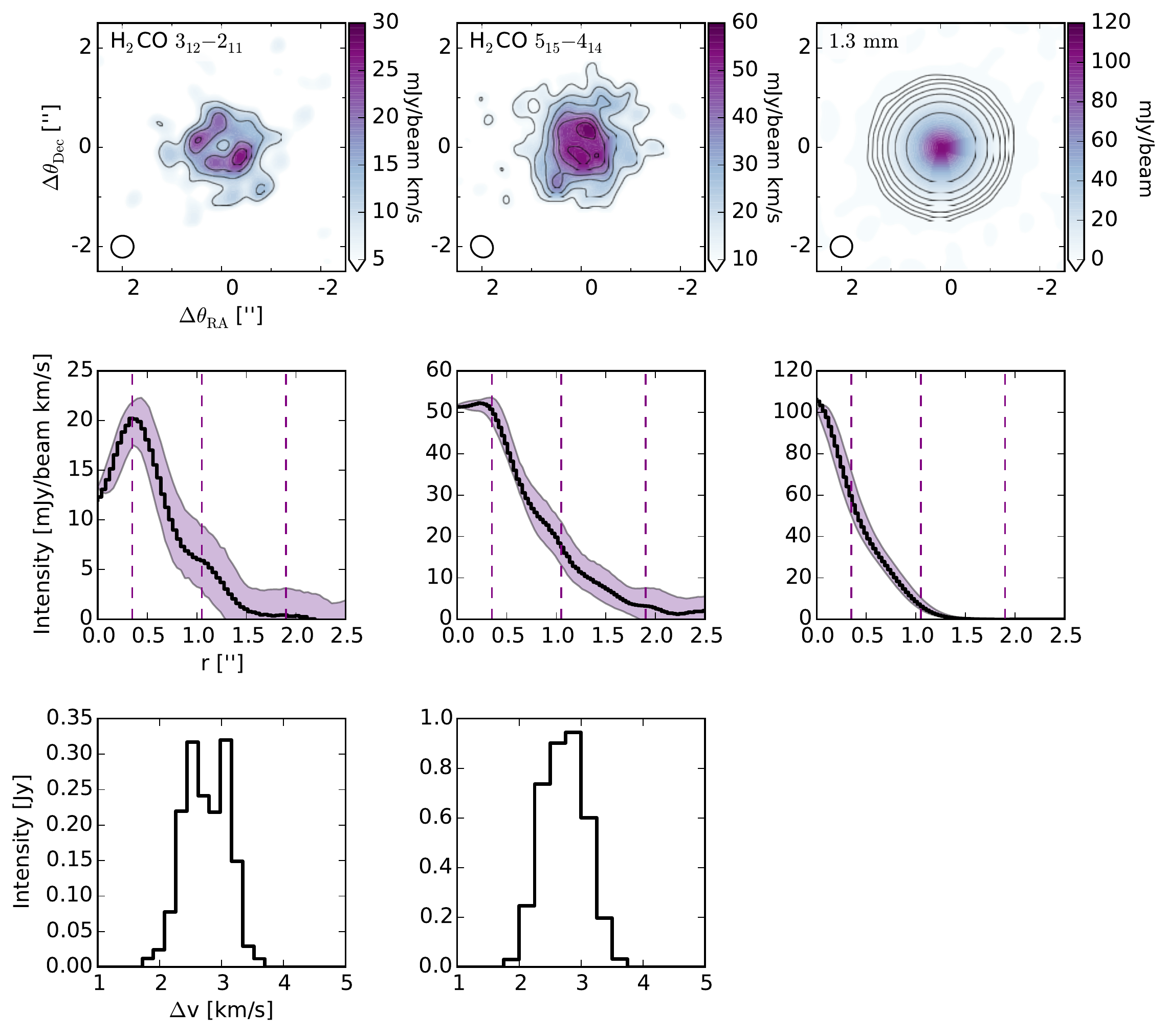}
\caption{Overview of observational results. {\it Top row:} Integrated emission maps of H$_2$CO $3_{12}-2_{11}$ (left), H$_2$CO $5_{15}-4_{14}$ (middle) and 1.3mm dust emission (right). The flux per beam is indicated by the color scales. The line contours are [3,5,7,9,11]$\sigma$ in the H$_2$CO images, and [4,8,16,32,64,128,256]$\sigma$  in the continuum image. {\it Middle row:} Radial profiles of the same H$_2$CO lines and dust continuum. The shaded region mark the 1$\sigma$ scatter in the intensity values. The three dashed lines mark the observed sub-structure in the H$_2$CO emission profiles. {\it Bottom row:} Extracted spectra using CLEAN masks. \label{fig:obs}}
\end{figure*}

Figure \ref{fig:obs} shows three different, more condensed visualizations of the H$_2$CO $3_{12}-2_{11}$ and H$_2$CO	$5_{15}-4_{14}$ data. The top row shows integrated emission or moment-zero maps of the H$_2$CO emission together with the 1.3~mm continuum. The images were generated in CASA using the {\it immoments} task without clipping, and include all channels with any emission above 3$\sigma$ ( 2.12-3.62 km/s for the $3_{12}-2_{11}$ line and 1.87-3.87 for the 
$5_{15}-4_{14}$ line). Notably, both H$_2$CO lines display central depressions, but the $3_{12}-2_{11}$ line depression is substantially deeper and appears consistent with a lack of emission at the source center. By contrast, the dust emission is centrally peaked at this spatial resolution. 

To decide whether the central emission depressions in H$_2$CO trace a lack of H$_2$CO toward the source center we have to exclude three other potential sources: continuum over-subtraction, line opacity, and dust opacity. To address the possibility of continuum over-subtraction, we applied the same continuum subtraction procedure to line free channels and imaged these channels identically to the H$_2$CO containing channels. We saw no significant emission hole in the resulting image. Second, we estimated the line opacity of both H$_2$CO lines using the toy models introduced below and find that they are optically thin throughout the disk for all considered abundance profiles. Finally, while we cannot exclude that dust opacity contribute some to the central H$_2$CO emission depression, it is unlikely to be major contributor. First, observations of other molecules, including CO isotopologues, at similar spatial resolution do not display an emission depression \citep{Schwarz16}. Second the depression is smaller for the higher frequency transition, where the dust opacity should be higher. We thus conclude that the central depressions in H$_2$CO emission reflect a real depletion in H$_2$CO abundance.

The different emission structures of H$_2$CO $3_{12}-2_{11}$, H$_2$CO $5_{15}-4_{14}$, and dust are further visualized in the middle row of Fig. \ref{fig:obs}, which displays azimuthally averaged radial profiles assuming an inclination of 7 degrees. In addition to the central hole, the $3_{12}-2_{11}$ data show a `bump' around $1\farcs05$ (62~au), and tentatively a second bump at $1\farcs9$ (110~au), similar to the location of structure in scattered light observations \citep{vanBoekel16}, indicative of a ringed H$_2$CO structure in the TW Hya disk. Based on recent ALMA observations, TW Hya hosts a series of dust rings between $0\farcs02$ and $\sim1\arcsec$ (1 and 59~au) \citep{Andrews16}. The observed H$_2$CO rings and sub-structure do not seem to correspond to any of the most pronounced dust gaps or peaks. The $1\farcs05$ bump appears to coincide with the edge of the millimeter dust disk, however. This is not the first time that chemical substructure has been observed at the edges of dust disks \citep{Oberg15,Huang16}, hinting at a real chemical change at the edges of large dust/pebble disks. Indeed, in the TW Hya disk, \citet{Schwarz16} found that there are CO isotopologue bumps at the same $1\farcs05$ disk location. This chemical change could be driven either by increased UV penetration \citep{Oberg15} or a temperature inversion \citep{Cleeves16a}.  The $5_{15}-4_{14}$ emission show similar, but less pronounced, radial structures compared to the H$_2$CO $3_{12}-2_{11}$ emission, and appears remarkably similar to the $5_{15}-4_{14}$ emission profile previously observed toward the DM Tau disk \citep{Loomis15}.

The third row of Fig \ref{fig:obs} shows the extracted spectra. For the spectra, the native spectral resolution was used rather than 0.25 km/s, which explains some of the different shapes of the two lines. The spectra were extracted from the spectral image cube using the CLEAN mask, and then summing up the emission in each channel. The resulting spectra provide a good measure of the total flux, but do not have any well-defined noise properties, and the total line fluxes and uncertainties listed in Table \ref{lines} are instead extracted from integrated flux maps without any clipping applied.

\section{H$_2$CO Toy Models}

There are multiple approaches in the literature for extracting information on molecular abundance profiles, including abundance retrieval using grids of parametric models \citep[e.g.]{Qi11,Oberg12,Qi13b,Oberg15} and Monte Carlo methods \citep{Teague15,Guzman17}, comparison between observed emission and astrochemistry disk model predictions \citep{Dutrey07,Teague15,Cleeves15}, and toy models \citep{Andrews12,Rosenfeld13}. Since we are in an exploratory phase for organic ice chemistry in disks, we adopt the latter approach in this study. Grid and MCMC methods by necessity rely on the assumption that the model being tested has the correct form, locking down the kind of model considered. In light of the wealth of substructure seen  in both the present H$_2$CO data and many other disks and molecules, it is not clear that we know what that form should be for individual disks and molecules. With that in mind, we present a series of toy models of increasing complexity to explore what families of H$_2$CO abundance structures are qualitatively consistent with the radial profiles and relative intensities of the observed H$_2$CO lines. We then compare these structures with previously published outputs of detailed astrochemistry codes in the next section.

In our model framework, H$_2$CO abundances are defined with respect to a pre-existing disk density and temperature model, developed to fit the TW Hya SED and the disk continuum emission \citep{Qi13b}. Briefly, the adopted TW Hya disk model is a steady viscous accretion disk, heated by irradiation from the central star and by accretion \citep{dAlessio99,dAlessio01,dAlessio06}. The disk model is axisymmetric, in vertical hydrostatic equilibrium, and the viscosity follows the $\alpha$ prescription \citep{Shakura76}. Energy is distributed through the disk by radiation, convection, and a turbulent energy flux. The penetration of the stellar and shock generated radiation is calculated, and takes into account scattering and absorption by dust grains. \citet{Qi13b} added a tapered exponential edge to the standard realization of this model framework to simulate viscous spreading \citep{Hartmann98,Hughes08, Qi11}. Following \cite{Qi11}, \citet{Qi13b} also modified the vertical temperature and density structure by changing the vertical distribution of large grains \citep{dAlessio06}. \citet{Qi13b} explored several different vertical dust grain distributions, of which we selected the intermediate case shown in the upper panels of Fig. \ref{fig:par}. It is important to note that despite decades of modeling, disk vertical temperature structures remain highly uncertain. We emphasize that until this uncertainty has been addressed, it is difficult to constrain the vertical emission layer of molecules in absolute terms, or derive accurate H$_2$CO abundances.  As shown below, we can, however, constrain important properties of the emitting layer without knowing the exact layer height. We also note that our model does not take into account the possible presence of a break in the thermal structure at the edge of the pebble disk \citep{Cleeves16a}, which could results in an underestimate of the temperature in the outer disk by 10\%--30\%. It is also worth noting that the model was constructed before the recent publication of a revised distance estimate to TW Hya \citep{GAIA16}, which somewhat affects the inferred disk physical parameters, but has a negligible impact on the conclusions of this study.

The physical disk model is populated with H$_2$CO using one of the parametric prescriptions described below.  The level populations of observed lines are computed using RADMC-3D version 0.39 \citep{Dullemond12}, assuming the gas is at local thermal equilibrium (LTE).  The critical densities of the $3_{12}-2_{11}$ and $5_{15}-4_{14}$  lines are $7\times10^5$ and  $2.6\times10^6$ cm$^{-3}$, respectively, at 20~K \citep{Shirley15}. Apart from the disk atmosphere, typical disk densities are above $1\times10^6$ cm$^{-3}$, justifying our assumption of LTE. We used the {\texttt{vis\_sample}\footnote{\texttt{the vis\_sample}  Python package  is publicly available at \url{$https://github.com/AstroChem/vis\_sample$} or in the Anaconda Cloud at \url{$https://anaconda.org/rloomis/vis\_sample$}} package to compute the Fourier Transform of the synthetic model and sample visibilities at the u -- v points of the observations. We finally integrate the emission and calculate the radial profiles using the same procedure as for observations. 

\begin{figure*}[ht!]
\figurenum{3}
\includegraphics[width=\linewidth]{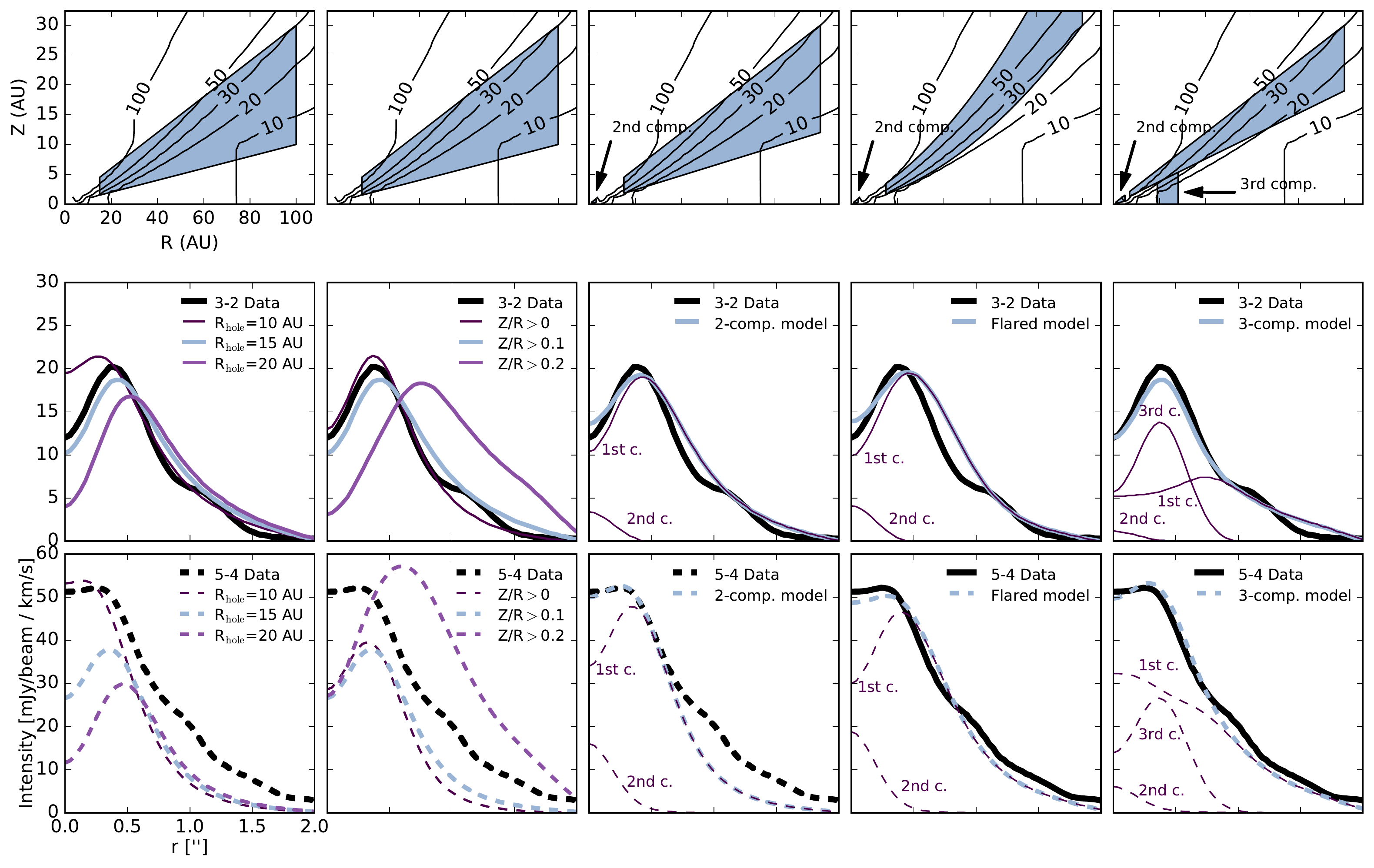}
\caption{Parametric toy model H$_2$CO distributions (top row) and the corresponding calculated $3_{12}-2_{11}$ and $5_{15}-4_{14}$ emission radial profiles (middle and bottom rows), shown together with observed profiles. In the toy model panel, the locations of H$_2$CO is marked by pale blue. Within each region the H$_2$CO abundance is either constant or set by a power law. The first two columns show the S1 model and the following three models the 2-component, flared and 3-component models. S2-S5 are not not shown explicitly in the top row. In the bottom two rows, the first column compares model outcomes with different inner holes (model S1-S3), the second column the effects of different vertical profiles (model S1, S4, S5), and  the third and fourth columns the improvements obtained by switching from a single component model to a two or three component model. Note the change in x-axis units from AU to arc seconds between the top row and the bottom two rows. All model parameters are listed in Table \ref{tab:models}, \label{fig:par}}
\end{figure*}

\begin{deluxetable*}{llcccccccc}
\tablecaption{H$_2$CO toy model parameters\label{tab:models}}
\tablehead{
\colhead{Parameter}   & \colhead{Description} & \colhead{S1}  & \colhead{S2}  & \colhead{S3}  & \colhead{S4}  & \colhead{S5}  & \colhead{2-comp.}  & \colhead{Flared}  & \colhead{3-comp.} 
}
\startdata
{\it Component 1:}\\
$x_{\rm 1AU}$	&abund. at 1~AU [10$^{-9}$ $n_{\rm H}$]	&$8$&$7$&10&0.3&500	&120		&0.2		&430\\
$\gamma$	&power law index	&-1.5	&-1.5	&-1.5	&-1.5&-1.5	&-2.0	&0.0		&-1.7\\
$r_{\rm in}$	&inner boundary [AU]	&15	&10	&20	&15	&15				&15		&15		&7\\
$r_{\rm out}$	&outer boundary [AU]	&100&100&100&100&100				&100		&100		&100\\
$(z/r)_{\rm low}$	&midplane boundary at $r_{\rm in}$ &0.1&0.1&0.1&0.0&0.2			&0.12	&0.11	&0.19\\
$(z/r)_{\rm up}$	&surface boundary  at $r_{\rm in}$  &0.3&0.3&0.3&0.3&0.3			&0.3		&0.23	&0.3\\
{\it Component 2:}\\
$x_{\rm c2}$	&inner disk abundance  [10$^{-9}$ $n_{\rm H}$]&--&--&--&--&--		&16		&20		&0.3\\
\multicolumn{3}{l}{\it Component 1 in flaring 2-comp.model:}\\
$(z/r)_{\rm low}$	&midplane boundary at $r_{\rm out}$&--&--&--&--&--		&--	&0.3	&--\\
$(z/r)_{\rm up}$	&surface boundary  at $r_{\rm out}$  &--&--&--&--&--		&--	&0.5	&--\\
{\it Component 3:}\\
$x_{\rm c3}$	&abundance [10$^{-9}$ $n_{\rm H}$] &--&--&--&--&--		&--		&--		&0.01\\
$r_{\rm c3, in}$	&inner  boundary [AU]	&--&--&--&--&--					&--		&--		&19\\
$r_{\rm c3, out}$&outer  boundary [AU]	&--&--&--&--&--					&--		&--		&28\\
$(z/r)_{\rm c3, low}$&lower boundary [AU]	&--&--&--&--&--				&--		&--		&0.0\\
$(z/r)_{\rm c3, up}$&upper boundary [AU]	&--&--&--&--&--				&--		&--		&0.19\\
\enddata
\end{deluxetable*}

\subsection{Single component models}

We begin by considering one of  the simplest possible distributions for H$_2$CO: a single abundance power law $x(r)= x_{\rm 1AU}\times r^\gamma$, confined by an inner and outer cut-off radius and a lower and upper boundary. The abundance $x$ is with respect to the total number of hydrogen nuclei per cm$^{-3}$, $r$ is the distance from the star in AU, and $\gamma$ a power law index. The lower and upper boundaries are defined by a constant $z/r$, where $z$ is the disk height above the midplane in AU. We fix the upper $(z/r)_{\rm up}$ and outer ($r_{\rm out}$) boundaries to 0.3 and 100~AU, respectively. We initially explored higher upper boundaries, and found that H$_2$CO  emission from more elevated disk regions was negligible as long as the lower boundary in term of $z/r$ is 0.2 or lower. The outer 100~AU boundary corresponds to the estimated outer edge of H$_2$CO emission in TW Hya. We then vary the inner and lower boundaries, as well as the power law parameters, to explore whether such a simple model can reproduce the observed H$_2$CO emission profiles, including the observed central depressions.  As shown in Fig. \ref{fig:par} (left two columns), a model with an inner radius of 15~AU and a lower layer boundary of 0--0.1 $z/r$ fits the shape of the observed $3_{12}-2_{11}$ radial profile quite well when adopting $\gamma=-1.5$. However, this model cannot reproduce the H$_2$CO $5_{15}-4_{14}$ radial profile flux level or shape. The predicted $5_{15}-4_{14}$ flux is too low at all radii, and the shape of the radial profile is wrong: compared to the observed $5_{15}-4_{14}$ emission the model depression toward the center is too deep, and in the outer disk the model profile falls off too steeply with radius.

To test whether other versions of the single-component power-law model could reproduce both the $3_{12}-2_{11}$ and $5_{15}-4_{14}$ emission, we set up a small grid of toy models varying the inner hole radius, the lower boundary of the emitting layer, and the power law index. For each model we select a inner radius and a lower layer boundary, and then adjust the power law coefficient and $x_{\rm 1AU}$ to obtain a reasonable `by-eye' fit to the H$_2$CO $3_{12}-2_{11}$ emission. The two left-hand columns of Fig. \ref{fig:par} show a sub-set of these models, focusing on the inner radius (first column), and the emitting layer location (second column). The model parameters are listed in Table \ref{tab:models} as S1-5 (where S stands for `single component model'). It is not possible to simultaneously reproduce the $3_{12}-2_{11}$ central hole, and the almost flat central profile of the $5_{15}-4_{14}$ emission with a single inner radius. The $5_{15}-4_{14}$ emission is also always under-predicted in the outer disk for the models that can reproduce the $3_{12}-2_{11}$ emission; that discrepancy increases with radius. This cannot be fixed by a uniform increase in the lower boundary since the required increase in $(z/r)_{\rm low}$ produces a H$_2$CO depression that extends farther out than is observed. It also cannot be explained by a radially dependent ortho-to-para ratio, since both lines are ortho lines. A single-component parametric model appears to be ruled out by the data. 

\subsection{Two- and three-component models}

First, consider the mismatch at small radii. Since the $3_{12}-2_{11}$ emission requires a considerable H$_2$CO abundance deficit and the $5_{15}-4_{14}$ emission profile requires some H$_2$CO on scales of 10~AU or less, it appears that the only way to reconcile the two is to add a hot unresolved H$_2$CO component that primarily contributes to the $5_{15}-4_{14}$ emission. We achieve this by adding a second H$_2$CO component between 1 and 3~AU, and between $z/r$ of 0.1 and 0.3 (Fig. \ref{fig:par}, third column). For simplicity, we set the H$_2$CO abundance to be constant in this layer. By adjusting the abundances in the hot component and the outer disk component, this model can reproduce the radial shapes and relative intensity levels of the $3_{12}-2_{11}$ and $5_{15}-4_{14}$ emission out to $\sim$0\farcs6  (Table \ref{tab:models}, 2-comp. model). Beyond this radius, the $5_{15}-4_{14}$ emission is always underproduced.

\begin{figure*}[ht!]
\figurenum{4}
\includegraphics[width=\linewidth]{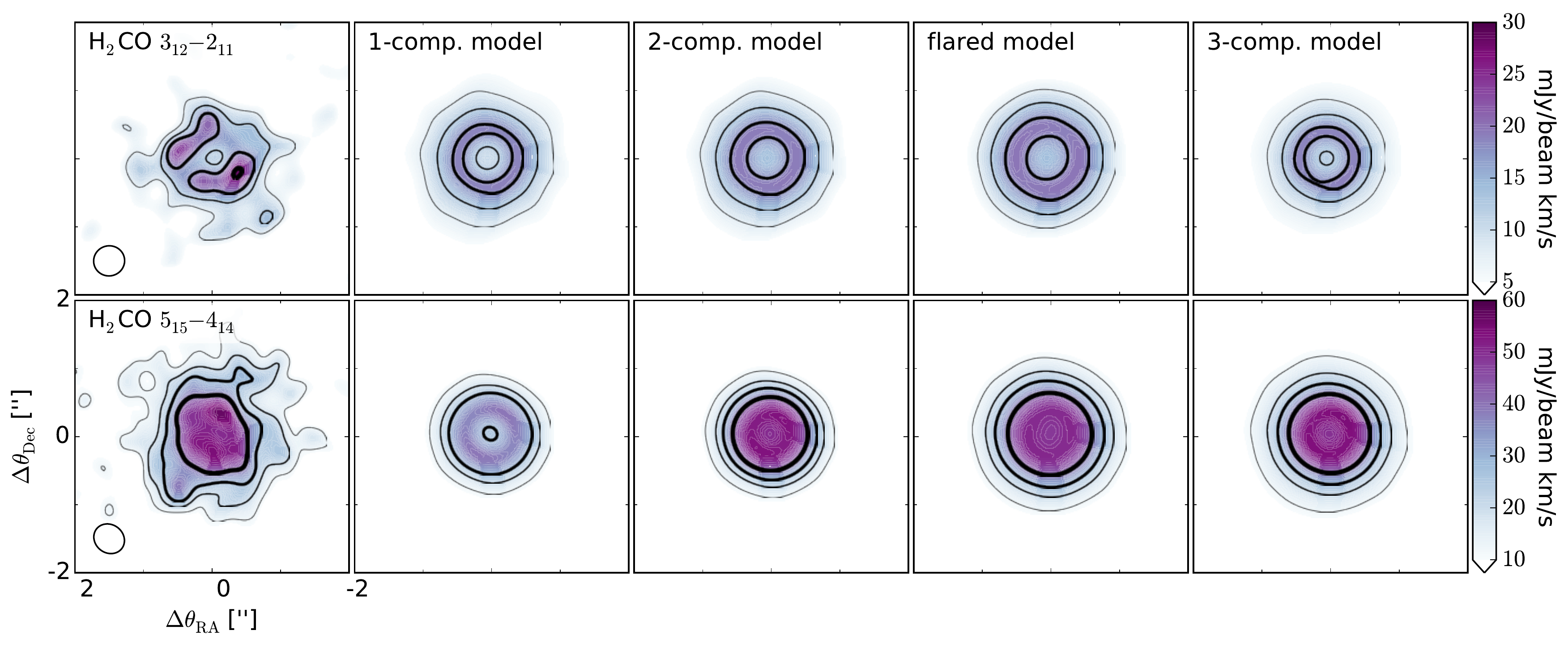}
\caption{Observed (first column) and simulated (column 2--5) H$_2$CO emission. The second column shows a single component power law model (S1 in Table \ref{tab:models}), the third column a 2-comp. model with a central hot component, the fourth column a flaring 2-comp. model and the fifth column a 3-comp. model, where the 3rd component constitutes a ring of emission between 19 and 28~AU. All model parameters are listed in Table \ref{tab:models}, \label{fig:mom0}}
\end{figure*}

Second, consider the mismatch at large radii. There is a radially increasing mismatch between predicted and observed $5_{15}-4_{14}$ emission for the models considered so far. This suggests that H$_2$CO is present in a warmer emitting region in the outer disk than what is achieved by constant $z/r$ boundaries. In the context of the adopted temperature profile, this implies that H$_2$CO is present in a more elevated layer (in units of $z/r$) in the outer disk compared to the emission peak at $\sim$0\farcs4. We consider two different parameterizations that could address this: a flaring H$_2$CO power law model (Fig. \ref{fig:par}, fourth column) and a three-component model (Fig. \ref{fig:par}, right-hand column). In the flaring H$_2$CO model, the lower and upper boundaries in units of $z/r$ increase linearly with radius between 15 and 100~AU. In the three component model, a midplane ring at 19--28~AU (3rd component) is combined with a power law abundance model at elevated, and therefore warm, disk layers ($z/r=0.19-0.3$) to produce excess $5_{15}-4_{14}$ emission in the outer disk. Both kinds of parameterizations can be set up to reproduce the overall radial shapes and relative emission levels of the $3_{12}-2_{11}$ and $5_{15}-4_{14}$ lines (Fig. \ref{fig:par}). The model parameters that best reproduce observations (by eye) are listed in Table \ref{tab:models}. 

\subsection{Model comparison}

The failure of the single power law models, and the simple 2-component model to reproduce the shape of the $5_{15}-4_{14}$ emission is also clearly seen in Fig. \ref{fig:mom0}, which shows integrated emission maps. As expected from the radial profile comparisons: the `optimized' flared H$_2$CO abundance model and the 3-comp. model reproduce the main features of the emission from both lines very well. It is important to note, however, that despite the close resemblance of observed and model emission, none of the toy models provide a good quantitative fit when comparing channel maps; there are significant residuals ($>5\sigma$). This is unsurprising, considering that these are toy models, but it emphasizes that the presented models should not be viewed as the final word on how H$_2$CO is distributed in the TW Hya disk. Rather, this exercise provides initial constraints on what families of models are consistent with the observations. 

In summary, the toy models  demonstrate that the $3_{12}-2_{11}$ emission requires a depleted H$_2$CO abundance in the inner disk, while the $5_{15}-4_{14}$ emission requires some H$_2$CO to  still be present there. This can be resolved if a H$_2$CO depletion is combined with a hot H$_2$CO component close to the star. The relative intensities of the two lines imply that H$_2$CO is vertically located close to the midplane (i.e. a lower boundary of 0-0.1 $z/r$) at intermediate disk radii,  beginning around 15~AU, and at more elevated disk layers at larger radii, beyond 60~au. In the context of the adopted temperature structure, this can be parameterized with either a flaring H$2$CO layer or with a 3-component model with constant $z/r$ vertical boundaries for each component.  None of these models reproduce the observed emission bump at 1$\farcs$05, which would require a fourth model component.

\section{Discussion}

\subsection{H$_2$CO abundance structure}

The H$_2$CO $3_{12}-2_{11}$ and $5_{15}-4_{14}$ line emission profiles in TW Hya belong to an increasingly large class of molecular lines that appear as emission rings in disks. Previous molecular emission rings have either been connected to dust deficiencies \citep[e.g.][]{Zhang14,Oberg15}, temperature or photon regulated gas-phase chemistry \citep[e.g.][]{Aikawa03,Qi08,Bergin16}, or snowlines \citep{Qi13b,Qi13c}.

In the case of H$_2$CO in TW Hya, we have demonstrated that the observed H$_2$CO emission profiles require an underlying abundance structure that fulfills the following conditions: 1) a distinct inner disk H$_2$CO component that is warm, 2) a second mid-to-outer disk component of H$_2$CO, where 3) the emitting layer resides at higher $z/r$ in the outer disk ($r>60$~AU) compared the to intermediate disk radii ($15<r<30$~AU). The last condition is a result of the presence of a high H$_2$CO $5_{15}-4_{14}$/$3_{12}-2_{11}$ intensity ratio in the outer disk. This implies a relatively warm H$_2$CO emitting layer in the outer disk, which in the adopted disk temperature model occurs at $z/r>0.2$.  

It is not clear, however, that the adopted temperature structure provides a good description of the TW Hya disk temperatures beyond the millimeter dust disk. \citet{Cleeves16a} found that the loss of millimeter grains in the outer disk will cause a temperature inversion, substantially increasing the temperature in the outer disk. There is observational  evidence that this takes place in other disks \citep{Huang16}, and it may well affect the TW Hya outer disk temperature profile. If a temperature inversion is present at the millimeter disk edge in TW Hya, two effects on the H$_2$CO emission may be expected: a change in the slope of all H$_2$CO intensity profiles  around the dust edge where the temperature inversion occurs, and excess $5_{15}-4_{14}$ emission everywhere in the outer disk, since the overall temperature will be elevated compared to standard disk modeling assumptions where the temperature monotonically decreases with radius. Both features are observed in the TW Hya disk, which is highly suggestive. Furthermore, \citet{Schwarz16} suggests that a bump in the CO emission at this radius may trace a second snowline, as would be expected if a temperature inversion is present. Better constraints on the outer disk temperature profile are needed before we can present a conclusive interpretation of the H$_2$CO outer disk emitting layer. In the meantime, we cannot tell whether the observed excess $5_{15}-4_{14}$ emission is due to an elevated emitting layer in the outer disk or excess temperatures in the outer disk compared to our adopted temperature structure.

In the inner disk, it is interesting to compare the location of the H$_2$CO inner radius and the CO snowline, since one of the proposed H$_2$CO formation pathways in disks is through CO ice hydrogenation. The CO snowline location is inferred from N$_2$H$^+$ observations to be at~30~AU \citep{Qi13c}. This is considerably outside of the 15--20~AU boundary where H$_2$CO first appears in the TW Hya disk (ignoring the inner hot component). However, this difference does not automatically exclude an icy origin of the outer disk H$_2$CO because the onset and completion of CO condensation can occur at different temperatures and therefore different disk locations. H$_2$CO ice formation is expected to become efficient at the onset of CO freeze-out, which is likely regulated by the temperature at which CO binds to H$_2$O ice. N$_2$H$^+$ gas-phase production becomes efficient when most CO has been depleted from the gas-phase, which is likely regulated by the lower CO:CO ice binding energy \citep{Collings03,Fayolle16}, since there should be sufficient CO in disks to form multi-layered CO ices in the outer disk. Furthermore, H$_2$CO formation on grains can begin at even higher grain temperatures than expected for CO freeze-out on water ice, since it only requires that some of the CO spends some of their time on grains. Based on these two considerations alone we would expect a substantial difference between the H$_2$CO inner radius and the CO snowline location, defined as the location where CO freeze-out nears completion.  In addition, a recent study shows that the inner radius of N$_2$H$^+$ rings only provides an upper limit to the CO snowline \citep{vantHoff16}. It is therefore possible that the CO snowline location in \citet{Qi13c} is overestimated by several AU. In light of these laboratory and theoretical results a large difference between the H$_2$CO  and the N$_2$H$^+$ inner edges is compatible with the proposed icy origin of H$_2$CO in the outer disk.

\subsection{H$_2$CO formation in the TW Hya disk}

H$_2$CO formation through gas and grain surface chemistry has been explored theoretically in a number of models \citep[e.g.][]{Aikawa03,Aikawa06,Willacy07,Willacy09,Walsh14}. Most recently, \citet{Loomis15} modeled the H$_2$CO abundance in a T Tauri disk with gas and grain surface formation, and with either gas or grain surface formation turned off. We focus our comparison on the three models that show the predicted H$_2$CO abundance structure in detail: \citet{Willacy09}, \citet{Walsh14}, and \citet{Loomis15}.

These three H$_2$CO model predictions share a number of important features. All contain a central H$_2$CO component, which is attributed to gas-phase formation of H$_2$CO. The extent of the inner component depends on the details of adopted  disk model, but is generally concentrated within 10~AU. All models also predict an outer disk H$_2$CO component at intermediate disk heights $0.2<z/r<0.4$. In \citet{Loomis15}, the inner and outer disk component connect, while in \citet{Willacy09} and \citet{Walsh14} the inner and outer disk components are distinct. This second component is explained by grain-surface formation of H$_2$CO. The component consists of a disk layer that is cold enough for CO to have a substantial residence time on grains, enabling hydrogenation to form H$_2$CO, and low-density enough for non-thermal desorption to maintain some of the formed H$_2$CO in the gas, either through release of chemical energy or photodesorption. 

In addition to these two components,  \citet{Willacy09} and \citet{Walsh14} predict a third, radially confined component close to the midplane at $\sim$20~AU. This component also appears to be due to grain surface formation followed by non-thermal desorption, and represents the midplane location where CO begins to reside for substantial amounts of time on grain surfaces. The lack of this component in \citet{Loomis15} may reflect different assumptions of radiation fields and desorption efficiencies in that model.

The observationally constrained H$_2$CO abundance structure in the TW Hya disk qualitatively agrees these model predictions. It is important to note, however, that non-thermal ice desorption efficiences are highly uncertain. As a result, the model predictions in the outer disk that rely on these pathways are order of magnitude estimates, at best.  Gas-phase chemistry will also continue to contribute to the overall H$_2$CO abundances throughout the disk through e.g. the very efficient CH$_3$+O reaction \citep{Atkinson06}. The relative importance of grain and gas-phase chemistry in the outer disk of TW Hya  is therefore difficult to quantify with certainty based on extracted column densities. The radial H$_2$CO abundance profiles provide stronger constraints, and the observed H$_2$CO emission profile in TW Hya is only expected if the grain-surface formation and desorption pathway produces the majority of the observed H$_2$CO exterior to 15~AU. Beyond that, it is not possible with the present data to distinguish between different model predictions, since the data are consistent both with a single flaring component in the outer disk (similar to \citet{Loomis15}, and a 3-component model, where the third component traces the midplane onset of H$_2$CO ice formation (similar to \citet{Willacy09} and \citet{Walsh14} ). Higher resolution data and a better constrained disk temperature profile used both for retrieval and chemistry model predictions are needed to resolve whether this third component is present in the TW Hya disk.

In either case, considering the good agreement between observational constraints on the H$_2$CO distribution in the TW Hya disk and astrochemical disk model predictions we can put some qualitative constraints on the H$_2$CO chemistry in the TW Hya disk. H$_2$CO forms close to the star due to warm gas-phase chemistry. If pebble drift is efficient, additional H$_2$CO gas may be delivered to the inner disk gas as the pebbles cross the H$_2$CO snowline \citep[e.g.][]{Oberg16b}. These processes enrich the gas in the terrestrial planet forming zone in gas-phase H$_2$CO. There is then a H$_2$CO chemistry desert until grain-surface chemistry kicks in around 15-20~AU through hydrogenation of CO ice, with some possible minor contribution from gas-phase chemistry. Further out in the disk, H$_2$CO ice continues to form at all disk heights where the temperature is sufficiently low, but H$_2$CO is only released efficiently into the gas-phase at intermediate to high disk layers, where non-thermal desorption is most efficient. 

In this scenario, we expect that CH$_3$OH would follow the H$_2$CO distribution, except it should lack the central component tracing gas-phase H$_2$CO formation chemistry. So far, only low-SNR observations of CH$_3$OH exist \citep{Walsh16}. The data favor a central cavity in the CH$_3$OH abundance  distribution, but the size of the cavity or the location of the emitting layer was not constrained. Higher SNR data are needed to provide a direct test of this prediction.

\section{Conclusions}

Using a series of toy models we demonstrated that to simultaneously reproduce the observed H$_2$CO $3_{12}-2_{11}$ and $5_{15}-4_{14}$ emission requires a distinct  hot H$_2$CO gas reservoir in the inner disk, and an extended H$_2$CO gas reservoir in the outer disk. The resulting two-component H$_2$CO model only reproduces the outer disk emission of both H$_2$CO lines if the H$_2$CO emitting layer increases in height (in terms of $z/r$) with radius, or if there is a hitherto undetected temperature inversion in the disk at the millimeter dust edge. The model approach thus informed us on what families of models that are consistent with the data. It is important to note that if only one of the two lines had been observed, the data would have been consistent with a single power law distribution, resulting in very different conclusions on the radial and vertical distribution of H$_2$CO in this disk. 

The inferred H$_2$CO structure is qualitatively consistent to what is predicted by astrochemical disk codes that includes both warm gas-phase chemistry and cold hydrogenation of CO on grains. This implies that similar to the much younger DM Tau system, the old TW Hya disk hosts an active gas-phase and grain-surface organic chemistry. This active disk chemistry should result in a time dependent organic composition in disks. It is thus likely that planetesimals assembling at different times during the lifetime of the disk will acquire different chemical compositions, and in particular different abundances of simple and complex organic molecules.

\acknowledgments

This paper makes use of the following ALMA data:
ADS/JAO.ALMA\#2013.1.00114.S  and ADS/JAO.ALMA\#2013.1.00198.S. ALMA is a partnership of ESO (representing
its member states), NSF (USA) and NINS (Japan), together with NRC
(Canada) and NSC and ASIAA (Taiwan), in cooperation with the Republic of
Chile. The Joint ALMA Observatory is operated by ESO, AUI/NRAO and NAOJ. The National Radio Astronomy Observatory is a facility of the National Science Foundation operated under cooperative agreement by Associated Universities, Inc. KI\"O also acknowledges funding from the Packard Foundation and an investigator award from Simons Collaboration on the Origins of Life (SCOL). VVG thanks support from the Chilean Government through the Becas Chile program. JH is supported by the National Science Foundation under Grant No. DGE-1144152. RL is supported by the National Science Foundation under Grant No. DGE-1144152. MRH is supported by a TOP grant from the Netherlands Organization for Scientific Research (NWO, 614.001.352).  CB is supported the European Research Council (ERC) under the European UnionÕs Horizon 2020 research and innovation programme (grant agreement No 638596)

\bibliographystyle{aasjournal}

\end{document}